\documentclass[preprints,article,accept,moreauthors,pdftex]{Definitions/mdpi}

\firstpage{1} 
\pubvolume{1}
\issuenum{1}
\articlenumber{0}
\pubyear{2022}
\copyrightyear{2022}
\datereceived{} 
\dateaccepted{} 
\datepublished{} 
\hreflink{https://doi.org/} 
\pdfoutput=1

\Title{Influence of $f$ Electrons on the Electronic Band Structure \\ 
of Rare-Earth Nickelates}

\TitleCitation{Influence of $f$ electrons on electronic band structure of rare-earth nickelates}

\Author{Andrzej Ptok $^{1,\ast}$\orcidA{}, Surajit Basak $^{1}$\orcidB{}, Przemys\l{}aw Piekarz $^{1}$\orcidC{}, and Andrzej M. Ole\'{s}$^{2,3}$\orcidD{}}

\AuthorNames{Andrzej Ptok, Surajit Basak, Przemys\l{}aw Piekarz, and Andrzej M. Ole\'{s}}

\AuthorCitation{Ptok, A.; Basak, S.; Piekarz, P.; Ole\'{s}, A. M.}

\address{%
$^{1}$ \quad Institute of Nuclear Physics, Polish Academy of Sciences, W. E. Radzikowskiego 152, \\ \hskip .5cm PL-31342 Krak\'{o}w, Poland \\
$^{2}$ \quad Max Planck Institute for Solid State Research, Heisenbergstrasse 1, D-70569 Stuttgart, Germany \\
$^{3}$ \quad Institute of Theoretical Physics, Jagiellonian University, Profesora Stanis\l{}awa \L{}ojasiewicza 11, \\ \hskip .5cm PL-30348 Krak\'{o}w, Poland}

\corres{Correspondence: aptok@mmj.pl}

\abstract{
Recently, superconductivity was discovered in the infinite layer of hole-doped nickelates NdNiO$_{2}$.
Contrary to this, superconductivity in LaNiO$_{2}$ is still under debate.
This indicates the crucial role played by the $f$ electrons on the electronic  structure and the pairing mechanism of infinite-layer nickelates.
Here we discuss the role of the electron correlations on the $f$ electron states and their influence on the electronic structure.
We show that the lattice parameters are in good agreement with the experimental values, independent of the chosen parameters within the DFT+$U$ approach.
Increasing Coulomb interaction $U$ tends to shift the $f$ states away from the Fermi level.
Surprisingly, independently of the position of $f$ states with respect to the Fermi energy, these states play an important role in the electronic band structure, which can be reflected in the modification of the NdNiO$_{2}$ effective models.
}


\graphicspath{{fig/}{./fig/}{.}}

\begin{document}


\section{Introduction}
\label{sec.intro}

Recently superconductivity was reported in hole-doped infinite-layer nickelate  NdNiO$_{2}$ at $9$-$15$~K, in thin-film samples grown on SrTiO$_{3}$~\cite{li.lee.19,zeng.tang.20,gu.li.20}.
Contrary to the thin layers, the bulk NdNiO$_{2}$ does not exhibit superconductivity~\cite{wang.zheng.20}.
Superconductivity was also reported in hole doped PrNiO$_{2}$~\cite{osada.wang.20,osada.wang.20b}, while its occurrence in LaNiO$_{2}$ it is still under debate~\cite{li.lee.19,osada.wang.21}.

These observations renewed interest in the pairing mechanism and the role played by the $f$ electrons in rare-earth nickelates~\cite{nomura.arita.22}. 
It was established recently that depending on the interactions in a two-band model for infinite-layer superconductors~\cite{Adh20}, pairing with $s$-wave or $d$-wave symmetry is possible \cite{Pli22}. In this context, several theoretical studies of NdNiO$_{2}$ were performed based on density-functional theory (DFT)~\cite{nomura.hirayama.19,jiang.si.19,Cho20,zhang.jin.20,wu.disante.20,zhang.lane.21,sakakibara.usui.20,wang.kang.20,botana.norman.20}, DFT+$U$~\cite{botana.norman.20,liu.ren.20,wan.ivanov.21,deng.jiang.21,gu.zhu.20}, or DFT+DMFT approaches
\cite{wang.kang.20,gu.zhu.20,lechermann.20,ryee.yoon.20,leonov.skornyakov.20,lechermann.20b,karp.botana.20,kitatani.si.20,karp.hampel.22,chen.jiang.22}.

The long range magnetic order was not reported experimentally (in achievable temperatures).
Nevertheless, recent experiments show the existence of magnetic correlations in LaNiO$_{2}$~\cite{ortiz.puphal.22} and magnetic and charge instabilities in NdNiO$_2$~\cite{lu.rossi.21,tam.choi.22,krieger.martinelli.22}.
The observed charge density wave (CDW), with the same in-plane wavevector (1/3,0) for Nd $5d$ and Ni $3d$ orbitals, disappears when superconductivity emerges in doped NdNiO$_{2}$~\cite{tam.choi.22}.
Similarly, for LaNiO$_{2}$, the CDW is characterized by incommensurate wavevector, and under doping the charge order diminishes and its wavevector shifts towards the commensurate order~\cite{rossi.osada.22}.
Such results suggest the existence of charge order and its potential interplay with antiferromagnetic fluctuations and superconductivity in infinite-layer nickelates.

The modern computational methods based on DFT require an approximate treatment of the electron exchange and correlation interactions.
Typically, the available exchange-correlation functional within the local density approximation (LDA) or generalized gradient approximation (GGA) works 
correctly for several different types of atoms. However, rare-earth metals with localized $f$ electrons require special treatment. Generally, the GGA functionals improve the obtained results compared to the LDA calculations~\mbox{\cite{delin.fast.98b,delin.fast.98}.} 
We remark that a good agreement with the experimental data is obtained 
only when self-interaction corrections or local Coulomb interactions 
between electrons are taken into account~\cite{strange.svane.99,soderlind.turchi.14,waller.piekarz.16}.

Good accuracy can be found within DFT+$U$ schemes, where the Hubbard interaction ($U_{eff} = U - J$, with on-site Coulomb interactions $U$ and on-site exchange Hund's interaction $J$) is treated in a mean-field manner, while obtained results can strongly depend on the used free parameters: $U$ and $J$~\cite{moor.harton.22}. 
In the literature, several sets of parameters in DFT+$U$ were used for Nd, for example, $U_{Nd} = 4.8$~eV and $J_{Nd} = 0.6$~eV estimated from constrained random-phase approximation~\cite{nilsson.sakuma.13}; $U_{Nd} = 6.76$~eV and $J_{Nd} = 0.76$~eV used for described Nd atom deposited on graphene~\cite{kozub.shick.16}; $U_{eff} = 6$~eV for study of the Nd cluster evolution~\cite{ma.zhang.21};
$U_{eff} = 7.5$~eV to reproduce experimental gap in neodymium gallate (NdGaO$_{3}$)~\cite{reshak.piasecki.09}. These values cover the expected range, while $U_{eff} = 10$~eV used for description of NdNiO$_{2}$~\cite{zhang.jin.20} we consider too large.
The range of used parameters raises the question of not only the role of $f$ electron on the electronic properties but also the realistic parameters describing correlation in rare-earth nickelates.

In this paper, we try to respond to this problem. 
Using the DFT(+$U$) calculation, we study the lattice parameters.
For different sets of $U$ and $J$, we discuss the electronic properties, 
focusing on the density of states and location of the $f$ electronic states.
The paper is organized as follows.
Details of the techniques used are present in Sec.~\ref{sec.method}.
Next, in Sec.~\ref{sec.res}, we present and discuss our theoretical results.
Finally, a brief summary is presented in Sec.~\ref{sec.sum}.


\begin{figure}[!b]
\centering
\includegraphics[width=\linewidth]{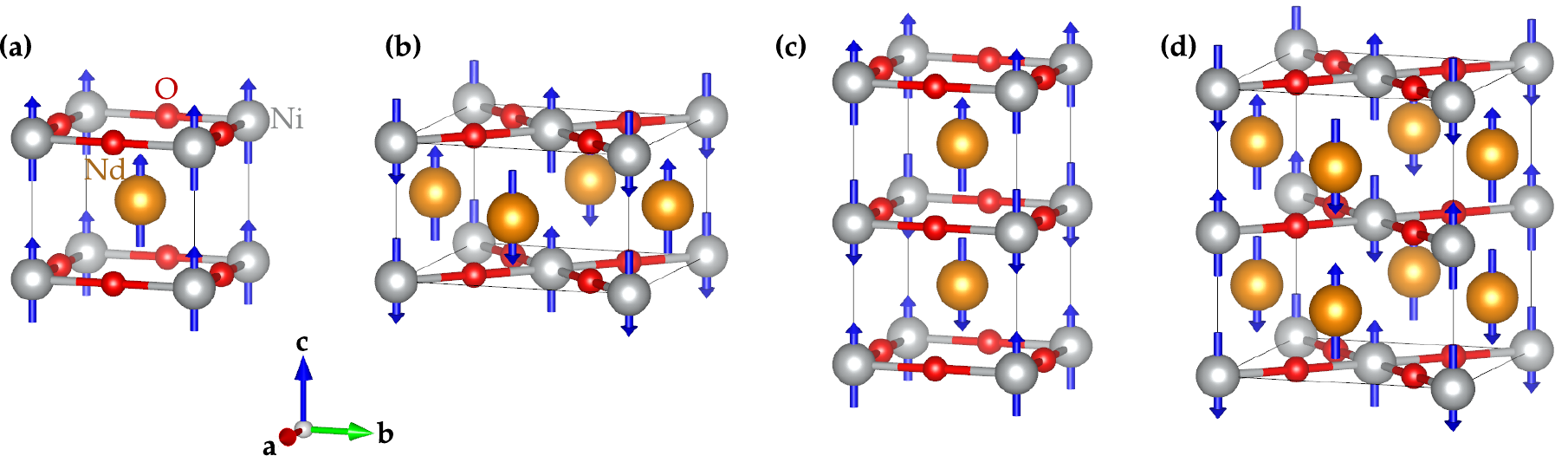}
\caption{
Phases of NdNiO$_{2}$ with magnetic order discussed in the paper: 
(a) ferromagnetic (FM); 
(b)~$C$-type antiferromagnetic ($C$-AFM); 
(c)~$A$-type antiferromagnetic ($A$-AFM); and 
(d) $G$-type antiferromagnetic ($G$-AFM). 
The magnetic states follow from the orientation of Ni-spins (at gray balls).
\label{fig.mag_crys}
}
\end{figure}

\section{Calculation details}
\label{sec.method}

The first-principles DFT calculations were performed using the projector augmented-wave (PAW) potentials~\cite{blochl.94} implemented in the {\it Vienna Ab initio Simulation Package} ({\sc Vasp}) code~\cite{kresse.hafner.94,kresse.furthmuller.96,kresse.joubert.99}. 
The exchange-correlation energy appearing in the Kohn-Sham equation was evaluated within the GGA, using the Perdew, Burke, and Ernzerhof (PBE) parameterization~\cite{pardew.burke.96}.
During calculations, we used the recommend pseudopotentials with the electronic configuration [He]$2s^{2}2p^{4}$ for O, [Ar]$3d^{8}4s^{2}$ for Ni, while for Nd configurations [Pd+$4f^{3}$]$5s^{2}5p^{6}5d^{1}6s^{2}$ or [Pd]$5s^{2}5p^{6}4f^{3}5d^{1}6s^{2}$, for the $f$ electrons treated as the core or the valence states, respectively.
Unless stated otherwise, the $f$ electrons were treated as valence electrons.
The energy cutoff was set to $800$~eV.
The correlation effects were introduced within DFT+$U$, proposed by Dudarev {et al.}~\cite{dudarev.botton.98}.
To avoid problems with calculation convergence, within the DFT+$U$, the spin--orbit coupling was not included.

In our study, we focused on the roles of the $f$ electrons (for different $U_{Nd}$ values).
To avoid too many free parameters, we set $U_{Ni} = 5.0$~eV, $J_{Ni} = 0.5$~eV, $J_{Nd} = 0.7$~eV, which are close to the  ones used in earlier studies~\cite{Cho20,liu.ren.20,botana.norman.20,been.lee.21,gao.peng.20,wan.ivanov.21,hampel.liu.19}.

In our investigation, we study systems with different types of magnetic order (see Fig.~\ref{fig.mag_crys}): non-magnetic (NM, not shown), ferromagnetic (FM), $A$-type antiferromagnetic \mbox{($A$-AFM)}, $C$-type antiferromagnetic ($C$-AFM), and $G$-type antiferromagnetic ($G$-AFM).
The NM and FM unit cells containing one formula unit were initially optimized with the $10 \times 10 \times 8$ {\bf k}--point grid.
For the $A$-AFM unit cell, built as a $1 \times 1 \times 2$ supercell and containing two formula units, $10 \times 10 \times 4$ {\bf k}-grid was used.
Similarly, the $C$-AFM unit cell is related to the $\sqrt{2}\times\sqrt{2}\times 1$ supercell containing two formula units---here the $7 \times 7 \times 8$ {\bf k}-grid was used.
Finally, the $G$-AFM unit cell (related to the $\sqrt{2}\times\sqrt{2}\times 2$ supercell and containing four formula units), was optimized with the $7 \times 7 \times 4$ {\bf k}-grid.
In all of the cases, the {\bf k}-grid in the Monkhorst--Pack scheme~\cite{monkhorst.pack.76} was used.
As the convergence condition of an optimization loop, we take the energy differences of $10^{-5}$~eV and $10^{-7}$~eV for ionic and electronic degrees of freedom, respectively.


\section{Results and discussion}
\label{sec.res}

\subsection{Crystal structure}

The bulk NdNiO$_{2}$ crystallizes in the P4/mmm symmetry (space group No. 123)~\cite{hayward.rosseinsky.03}.
The atoms are located in the high symmetry Wyckoff positions: Ni $1a$(0,0,0), O $1d$(0,1/2,0), and Nd $2f$(1/2,1/2,1/2).
Experimentally, the average values of lattice constants are \mbox{$a=3.920$~\AA} and $c = 3.275$~\AA\ (the lattice constants obtained by minimization of magnetic structures for increasing values of $U_{Nd}$ are collected in Table~\ref{tab.param})~\cite{hayward.rosseinsky.03,li.he.20,lin.gawryluk.22}.


\begin{table}[!t]
\caption{
\label{tab.param}
Comparison of the predicted NdNiO$_{2}$ system parameters.
}
\begin{tabular}{lrcccc}
Phase $\quad$ & \multicolumn{2}{c}{Lattice constant (\AA)} & $\quad E$ (eV/f.u.)  $\quad$ & $\mu_\text{Nd}$ ($\mu_{B}$) & $\mu_\text{Ni}$ ($\mu_{B}$) \\
 & a & c & & & \\
\hline 
\multicolumn{6}{c}{Experimental values} \\
\hline
Ref.~\cite{hayward.rosseinsky.03} & 3.919 & 3.307 & $-$ & $-$ & $-$ \\
Ref.~\cite{li.he.20} & 3.914 & 3.239 & $-$ & $-$ & $-$ \\
Ref.~\cite{lin.gawryluk.22} & 3.928 & 3.279 & $-$ & $-$ & $-$ \\
\hline 
\multicolumn{6}{c}{DFT (GGA PBE)} \\
\hline
NM & 3.909 & 3.314 & $-$28.464 & $-$ & $-$ \\
FM & 3.896 & 3.277 & {\bf $-$31.462} & 3.155 & 0.184 \\
$A$-AFM & 3.894 & 3.312 & $-$31.397 & 3.123 & 0.081 \\
$C$-AFM & 3.900 & 3.267 & $-$31.448 & 3.123 & 0.404 \\
$G$-AFM & 3.903 & 3.283 & $-$31.424 & 3.083 & 0.406 \\
\hline 
\multicolumn{6}{c}{DFT+$U$ (GGA PBE, $\quad$ $U_{Ni}=5.0$~eV, $J_{Ni} = 0.5$~eV)} \\
\hline
NM & 3.891 & 3.306 & $-$26.203 & $-$ & $-$  \\
FM & 3.925 & 3.249 & $-$29.618 & 3.129 & 0.854 \\
$A$-AFM & 3.926 & 3.273 & $-$29.568 & 3.089 & 0.892 \\
$C$-AFM & 3.952 & 3.216 & {\bf $-$29.594} & 3.076 & 1.007 \\
$G$-AFM & 3.871 & 3.303 & $-$29.170 & 3.112 & 0.000 \\
\hline 
\multicolumn{6}{c}{DFT+$U$ (GGA PBE, $\quad$ $U_{Ni} = 5.0$~eV, $J_{Ni} = 0.5$~eV, $\quad$  $U_{Nd} = 2.0$~eV, $J_{Nd} = 0.7$~eV)} \\
\hline
NM & 3.891 & 3.306 & $-$26.203 & $-$ & $-$  \\
FM & 3.937 & 3.292 & $-$29.287 & 3.125 & 0.924 \\
$A$-AFM & 3.929 & 3.307 & $-$28.948 & 3.138 & 0.867 \\
$C$-AFM & 3.941 & 3.263 & {\bf $-$29.303} & 3.046 & 0.962 \\
$G$-AFM & 3.879 & 3.333 & $-$28.629 & 3.162 & 0.000 \\
\hline 
\multicolumn{6}{c}{DFT+$U$ (GGA PBE, $\quad$ $U_{Ni} = 5.0$~eV, $J_{Ni} = 0.5$~eV, $\quad$  $U_{Nd} = 4.0$~eV, $J_{Nd} = 0.7$~eV)} \\
\hline
NM & 3.891 & 3.306 & $-$26.203 & $-$ & $-$ \\
FM & 3.940 & 3.284 & $-$28.712 & 3.061 & 0.893 \\
$A$-AFM & 3.940 & 3.290 & $-$28.714 & 3.056 & 0.938 \\
$C$-AFM & 3.950 & 3.263 & {\bf $-$28.742} & 3.031 & 0.940 \\
$G$-AFM & 3.949 & 3.264 & $-$28.740 & 3.004 & 0.945 \\
\hline 
\multicolumn{6}{c}{DFT+$U$ (GGA PBE, $\quad$ $U_{Ni} = 5.0$~eV, $J_{Ni} = 0.5$~eV, $\quad$  $U_{Nd} = 6.0$~eV, $J_{Nd} = 0.7$~eV)} \\
\hline
NM & 3.891 & 3.306 & $-$26.203 & $-$ & $-$ \\
FM & 3.943 & 3.286 & $-$28.483 & 3.016 & 0.905 \\
$A$-AFM & 3.944 & 3.287 & $-$28.483 & 3.006 & 0.936 \\
$C$-AFM & 3.955 & 3.274 & {\bf $-$28.517} & 3.017 & 0.941 \\
$G$-AFM & 3.954 & 3.275 & $-$28.515 & 2.997 & 0.948 \\
\hline 
\multicolumn{6}{c}{DFT+$U$ (GGA PBE, $\quad$ $U_{Ni} = 5.0$~eV, $J_{Ni} = 0.5$~eV, $\quad$ $U_{Nd} = 8.0$~eV, $J_{Nd} = 0.7$~eV)} \\
\hline
NM & 3.891 & 3.306 & $-$26.203 & $-$  & $-$  \\
FM & 3.950 & 3.294 & $-$28.321 & 3.014 & 0.910 \\
$A$-AFM & 3.948 & 3.297 & $-$28.319 & 3.004 & 0.933 \\
$C$-AFM & 3.958 & 3.285 & {\bf $-$28.351} & 3.013 & 0.944 \\
$G$-AFM & 3.960 & 3.286 & $-$28.349 & 2.996 & 0.951 \\
\hline 
\multicolumn{6}{c}{DFT+$U$ (GGA PBE, $\quad$ $U_{Ni} = 5.0$~eV, $J_{Ni} = 0.5$~eV, $\quad$ $U_{Nd} = 9.0$~eV, $J_{Nd} = 0.7$~eV)} \\
\hline
NM & 3.891 & 3.306 & $-$26.203 & $-$ & $-$ \\
FM & 3.951 & 3.299 & $-$28.256 & 3.016 & 0.911 \\
$A$-AFM & 3.951 & 3.301 & $-$28.255 & 3.006 & 0.933 \\
$C$-AFM & 3.959 & 3.288 & {\bf $-$28.285} & 3.014 & 0.946 \\
$G$-AFM & 3.962 & 3.289 & $-$28.283 & 2.998 & 0.953 \\
\hline 
\end{tabular}
\end{table}


As we can see, the obtained lattice parameters $a$ and $c$ agree pretty well with the experimental results, independent of the values of $U$ and $J$ parameters taken at Ni and Nd ions, respectively.
In this context, the crystal structure cannot be used as an argument to set some specific values of $U$ and $J$ within DFT calculations.

\begin{figure}[!t]
\begin{adjustwidth}{-\extralength}{0cm}
\centering
\includegraphics[width=\linewidth]{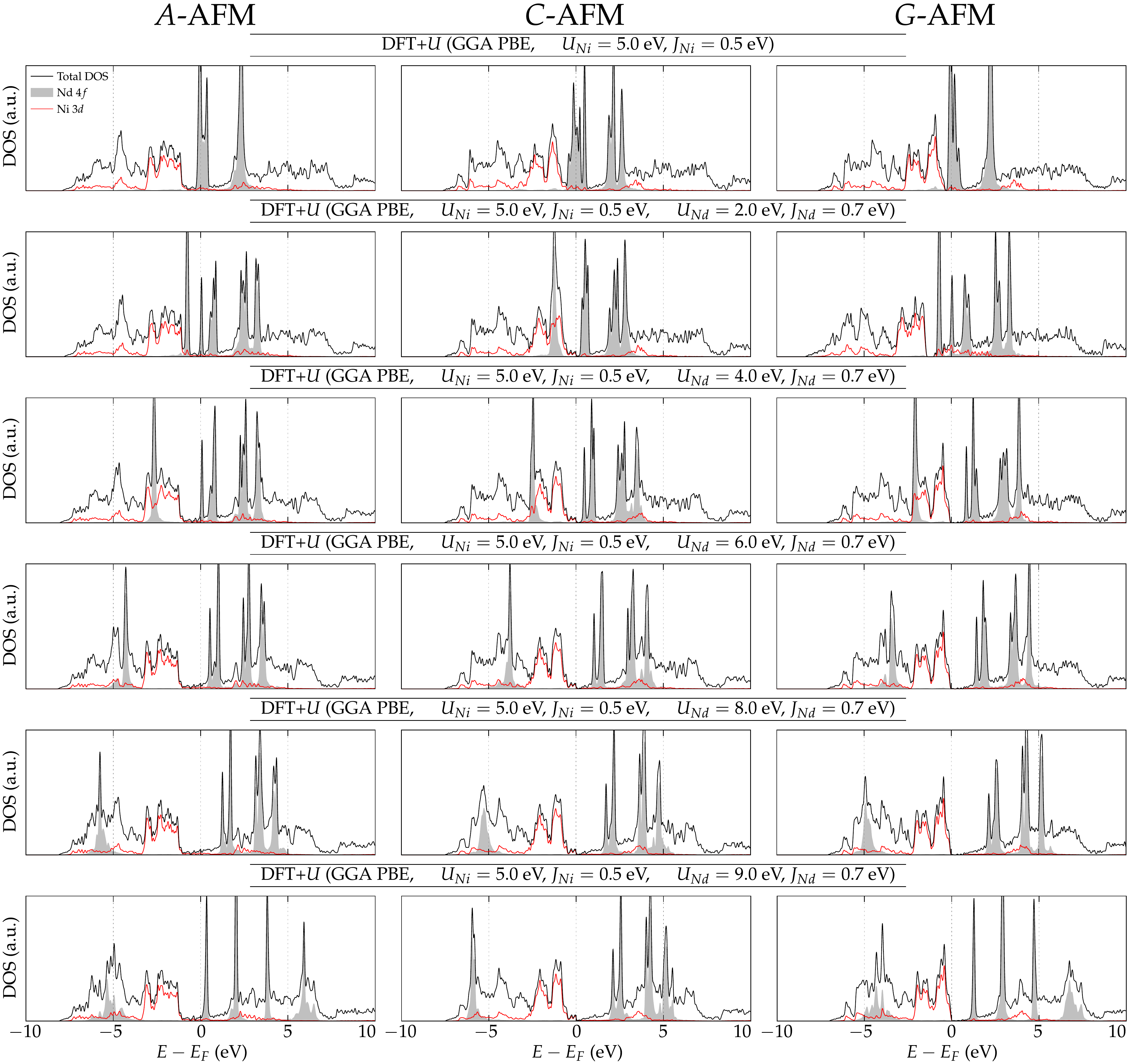}
\end{adjustwidth}
\caption{
Comparison of the electronic density of states (DOS) of NdNiO$_{2}$ for different magnetic configurations and model parameters. 
We take constant values of $U_{Ni} = 5.0$ eV and $J_{Ni} = 0.5$ eV; the Coulomb parameter at Nd ions $U_{Nd}$ increases as marked, while Hund's exchange is constant and stabilizes locally high-spin states, $J_{Nd} = 0.7$ eV.
The total DOS is shown by black line, Ni $5d$ states by red line, while multiplet structure of the Nd $4f$ electrons are indicated by gray-shaded maxima.
}
\label{fig.dos}
\end{figure}


\subsection{Magnetic ground state}

The energies of different magnetic orders are also collected in Table~\ref{tab.param}.
For each configuration, the magnetic ground state is marked by a bold number in the energy column. 
As we can see, for the calculation without the local interactions within the GGA PBE scheme, the magnetic ground state is incorrectly predicted as having FM order.
However, when Coulomb interactions are included in a more realistic DFT+$U$ picture, the magnetic ground state is always given as $C$-AFM phase (which is in agreement with the previous study~\cite{kapeghian.botana.20}).
In Fig. 2 we present the DOS for magnetic states up to the case with strong Coulomb interaction $U_{Nd} = 9.0$~eV and $J_{Nd} = 0.7$~eV, which we take as an upper limit of $U_{Nd}$. 
Close to $U_{Nd}=9$ eV, the energies of $C$-AFM and $G$-AFM phases are nearly degenerate.

For all the assumed parameters, the magnetic moments of Nd (from $f$ electronic states) appear to be close to $3$~$\mu_{B}$ but the dependence on $U_{Nd}$ is weak.
Contrary to this, the Ni magnetic moments strongly depend on the assumed magnetic order and on the used Kanamori parameters to describe the Coulomb interactions ($U$ and $J$) at Ni ions. 
For example, for weak interactions (i.e., small values of $U_{Nd}$), the Ni magnetic moment is equal to zero for $G$-AFM configuration (see Table 1).
Increasing $U_{Ni}$ leads to the stabilization of Ni magnetic moments, stemming from one hole states, around $1$~$\mu_{B}$.
The obtained Ni magnetic moments are comparable to the ones previously reported for NdNiO$_{2}$~\cite{liu.ren.20}, and mostly twice larger than observed in LaNiO$_{2}$~\cite{lee.pickett.04} (for similar Hubbard-like parameters assumed on Ni ions).

The energies of $C$-AFM and $G$-AFM are comparable (with small differences between the ground states). 
The final ground state energy depends on many parameters, and any strong statement cannot be given.
Nevertheless, the magnetic moment lead to the more energetically favorable magnetic ordered state (with AFM order).
As we mention in the introduction, the long range magnetic order was not reported in NdNiO$_{2}$. 
However, experimental results suggest realization of strong anitferromagnetic fluctuations, which can be reflected in magnetic ground state found within DFT+$U$ manner.

\subsection{Electronic density of states}

For used sets of $U$ and $J$ parameters, we calculated the electronic density of states (see Fig.~\ref{fig.dos}).
In the absence of a correlation effect on Nd atoms (first row), the $f$ electronic states (marked by gray-filled areas) are located in close vicinity of the Fermi level. Having finite $J_{Nd}$ stabilizes the high-spin states at Nd ions and gives the multiplet excited states shown in Fig.~\ref{fig.dos}.
In the presence of non-zero $U_{Nd}$ and $J_{Nd}$, the $f$ energy levels are split into several peaks.
In such cases, the increasing $U_{Nd}$ leads to shifting the $f$ states peaks from the vicinity of the Fermi level (cf. panels from top to bottom).
This is true for both (occupied and unoccupied) $f$ electron states.
This trend is not preserved only for ``extremely'' large values of Coulomb
interaction $U$, i.e., $U_{Nd} \gg 8.0$~eV (left and right panels on the lowest row in Fig.~\ref{fig.dos}).
In this case, the lowest band of the unoccupied $f$ states is separated and shifted closer to the Fermi energy.

\begin{figure}[!b]
\begin{adjustwidth}{-\extralength}{0cm}
\centering
\includegraphics[width=\linewidth]{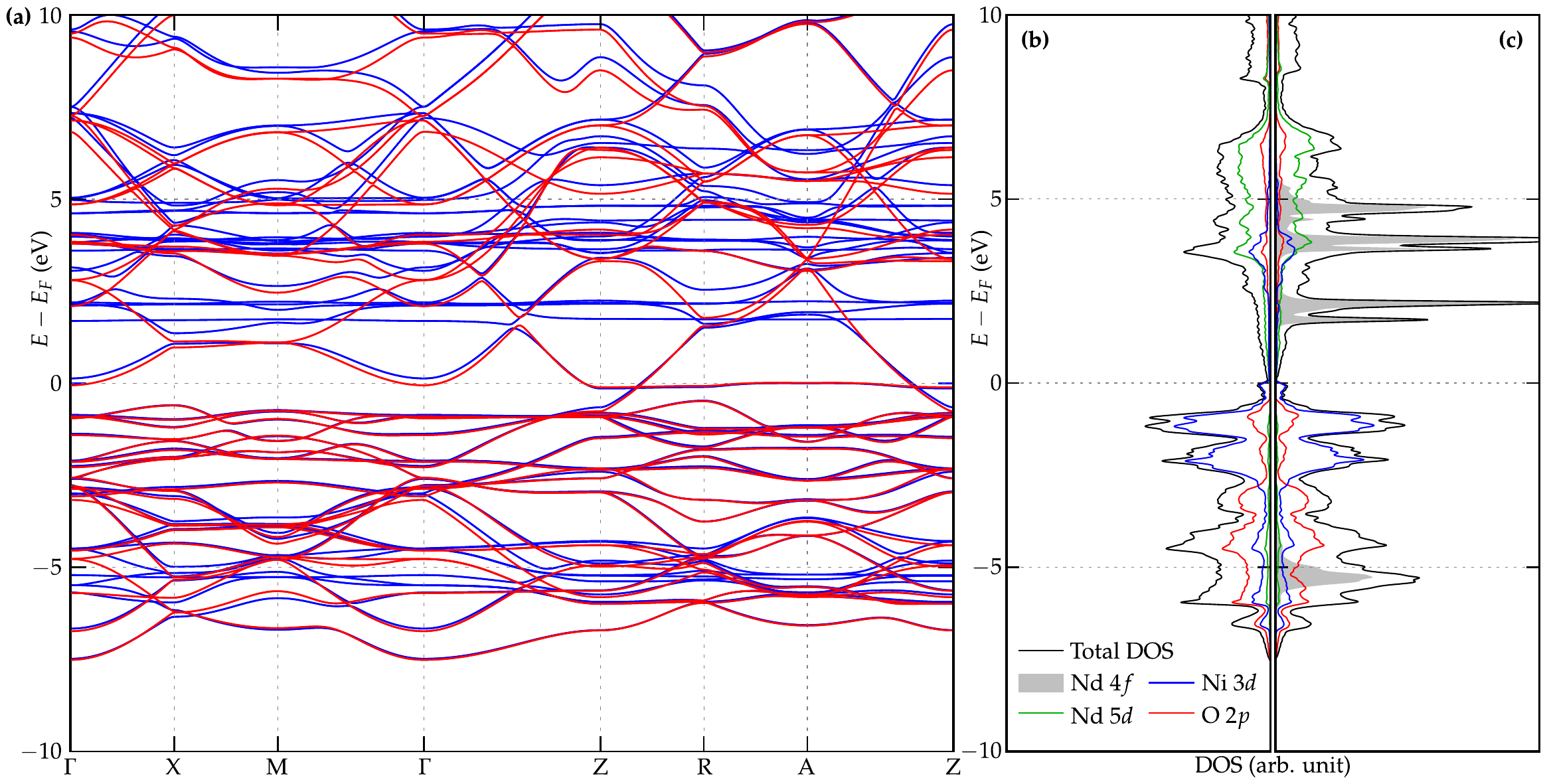}
\end{adjustwidth}
\caption{
Electronic band structure (a) and density of states (panels (b) and (c)) of NdNiO$_{2}$ with $C$-AFM spin order obtained for different treatments of the Nd $f$ electrons.
The band structure (a) for the $f$ electrons treated as core or as valence states, are presented by red and blue lines, respectively.
Related density of states, for the $f$ electrons treated as core or as valence states, are presented in panels (b) and (c), respectively.
Results obtained within DFT+$U$ (GGA PBE method) for the Coulomb interactions $U_{Ni} = 5.0$~eV, $J_{Ni} = 0.5$~eV, $U_{Nd} = 8.0$~eV, $J_{Nd} = 0.7$~eV.
}
\label{fig.band}
\end{figure}

\subsection{The changes in the electronic bands due to $f$ electron states}

Now we briefly discuss the role played by $f$ electron states in the electronic band structure (Fig.~\ref{fig.band}).
To describe their influence, we calculate the electronic band structure with $f$ electrons treated as valence states, as well as core states (results are shown as blue and red lines in Fig.~\ref{fig.band}, respectively). 
Here, we take the magnetic ground state ($C$-AFM) obtained for $U_{Nd} = 8.0$~eV and $J_{Nd} = 0.7$~eV.

In the case of $f$ states treated as valency electrons, the occupied $f$ states are located at energies $\mathcal{E} \in (-6.0,-4.0)$~eV.
Unoccupied $f$ states are located closer to the Fermi level---in the electronic DOS, the peaks of $f$ states appear around $1.7$, $2.2$, $3.6$, $3.9$, and $4.7$~eV.
The $f$ states are visible in the band structure in form of nearly flat bands.
The hybridization between $d$ and $f$ states leads to renormalization of the electronic band structure (around positions of $f$ states) with respect to the band structure with $f$ electrons treated as the core ones (cf. blue and red lines in Fig.~\ref{fig.band}(a)).

For the electronic states below the Fermi level, the $d$-$f$ hybridization does not modify the band structure strongly---the band renormalization is relatively weak. 
However, the situation is much more complicated in the case of the unoccupied states located in the close vicinity of the Fermi level.
Introduction of the $f$ electrons leads here to the modification of the band structure---the bottom of the band at $\Gamma$ point is shifted above the Fermi level.
This can lead to the modification of the Fermi surface, and to the disappearance of a Fermi pocket.
The forms of unoccupied band structures are similar, while the band structure shift is clearly visible---the bands for the calculation with $f$ electrons as the valence one are shifted to higher energies.

The band structure modification with the introduction of the $f$ electrons to the calculations is reflected in the electronic DOS (see Fig.~\ref{fig.band}(b) and (c), for DOS obtained when the $f$ electrons are treated as core or valence states, respectively).
As we can see, for the occupied states, the main peaks in the DOS structure, are located mostly at the same energies while total DOS is modified only by additional $f$ states.
Contrary to this, for unoccupied states, the shift of the states to higher energies is well visible (cf. electronic DOS for the case when the $f$ orbitals are treated as a valence or core states, presented on Fig.~\ref{fig.band}(b) and (c), respectively).
Largest differences are visible for Nd $5d$ states.
However, some modifications are also visible for Ni $3d$ and O $2p$ states.
Such effects should be included in further descriptions of the strong hybridisation between Ni $3d$ and Nd $5d$ states observed experimentally~\cite{tam.choi.22}.

As we mentioned earlier, the introduction of the $f$ states in the calculations as a valence electrons can lead to the modification of the Fermi surface (by disappearance of the pocket at $\Gamma$ point).
The band structure renormalization should be reflected in the effective model describing the NdNiO$_{2}$ system.
Recently many models were developed, from one band model based on Ni $d_{x^{2}-y^{2}}$ orbitals~\cite{kitatani.si.20,lee.pickett.04}, to the multiband models, like: 
(i) two orbital models 
(e.g. based on Ni $d_{x^{2}-y^{2}}$ and $d_{z^{2}}$ orbitals~\cite{sakakibara.usui.20}; 
Nd $d_{z^{2}}$ and Ni $d_{xy}$ orbitals~\cite{been.lee.21}; 
Ni $d_{x^{2}-y^{2}}$ and extended $s$-like state~\cite{Pli22}), 
(ii) three orbital models 
(containing Ni $d_{xy}$ and $d_{z^{2}}$ orbitals with additional interstitial $s$ orbital~\cite{gu.zhu.20}; 
Ni $d_{x^{2}-y^{2}}$ and Nd $d_{z^{2}}$ orbitals, with additional interstitial orbital mixing $s$ and $d_{xy}$ orbitals centered on Nd ions~\cite{nomura.hirayama.19}; 
effective Nd $d_{z^{2}}$, Nd $d_{x^{2}-y^{2}}$, and Ni $d_{xy}$ orbitals~\cite{liu.ren.20}; 
Ni $d_{z^{2}}$ and $d_{x^{2}-y^{2}}$ orbitals, and self-doped $s$-like orbital~\cite{lechermann.20b}),
or (iii) full 13-orbitals tight binding model in the Wannier orbital basis~\cite{karp.hampel.21}.
Nevertheless, the role of $f$ orbitals should be introduced indirectly to the model, by the effective modification of tight binding hopping parameters.


\section{Summary and conclusions}
\label{sec.sum}

For the suitable parameters of the local electron Coulomb interactions ($U$ and $J$ parameters) within the DFT+$U$ scheme, NdNiO$_{2}$ possesses the AFM ground state order. 
The calculations predict $C$-AFM ground state order which we treat here as a prediction for future experimental studies.
In the absence of Coulomb interactions at Nd atoms, the $f$ states are located in close vicinity of the Fermi level. 
Introduction of Coulomb interaction $U_{Nd}$ at Nd atoms, within the DFT+$U$ manner, generates the shift of the $f$ states, and then the energies of these states are far away from the Fermi level.

In all cases, the $f$ states induce large magnetic moments at Nd ions and by this change of the electronic structure, they still play an important role.
Even when the $f$ states are located above the Fermi level, the $d$-$f$ hybridization leads to the modification of the electronic band structure.
In particular, the shape of the Fermi surface is modified by $d$-$f$ hybridization which strongly impacts the pairing susceptibility.
We show that this effect can strongly affect the electronic band structure, which is crucial for the adequate theoretical description and understanding of the physical properties of superconducting nickelates.

\vspace{6pt} 

\authorcontributions{
A.P. initialized this project; A.P., S.B., P.P. performed theoretical calculations; A.M.O. reviewed the cited literature; A.P. prepared the first version of the manuscript. 
All authors contributed to the discussion and analysis of the results of the manuscript. Furthermore,
all authors have read and agreed to the published version of the manuscript.
}

\funding{
We kindly acknowledge support from National Science Centre (NCN, Poland) under Project No. 2021/43/B/ST3/02166.  
}

\institutionalreview{Not applicable}

\informedconsent{Not applicable}

\dataavailability{Not applicable} 

\acknowledgments{
We acknowledge insightful discussions with Wojciech Brzezicki.
Some figures in this work were rendered using {\sc Vesta}~software~\cite{momma.izumi.11}.
A.P. appreciates funding in the framework of scholarships of the Minister of Science and Higher Education (Poland) for outstanding young scientists (2019 edition, No. 818/STYP/14/2019).
}

\conflictsofinterest{The authors declare no conflict of interest.} 

\begin{adjustwidth}{-\extralength}{0cm}

\reftitle{References}

\bibliography{biblio.bib}

\end{adjustwidth}
\end{document}